\documentclass[reprint,amsmath,amssymb,aps]{revtex4-1}
\usepackage{physics}
\usepackage{graphicx}
\usepackage{dcolumn}
\usepackage{bm}
\usepackage{amssymb}
\usepackage{braket}
\usepackage{mathtools}
\usepackage{comment}
\usepackage{epstopdf}

\begin{document}

\graphicspath{ {./images/} } 
\title{Complex refraction metasurfaces for locally enhanced propagation through opaque media}
\author{Sinuh\'e Perea-Puente}
\author{Francisco J. Rodr\'iguez-Fortu\~no}%
 \email{Corresponding author: francisco.rodriguez-fortuno@kcl.ac.uk.}

\affiliation{
Department of Physics, King's College London, Strand, London WC2R 2LS, United Kingdom\\}

\date{\today}

\begin{abstract}
Metasurfaces with linear phase gradients can redirect light beams. We propose controlling both phase and amplitude of a metasurface to extend Snell's law to the realm of complex angles, enabling a non-decaying transmission through opaque media with complex refractive indices. This leads to the discovery of non-diffracting and non-decaying solutions to the wave equation in opaque media, in the form of generalised cosine and Bessel-beams with a complex argument. While these solutions present nonphysical exponentially growing side tails, we address this via a windowing process, removing the side tails of the field profile while preserving significant transmission enhancement through an opaque slab on a small localized region. Such refined beam profiles may be synthesized by passive metasurfaces with phase and amplitude control at the opaque material's interface. Our findings, derived from rigorous solutions of the wave equation, promise new insights and enhanced control of light propagation in opaque media.

\end{abstract}

\maketitle

\section{Introduction}
Propagation through a medium that would normally block, absorb, interfere or distort the passage of incident electromagnetic waves is a sought-after phenomenon, with the intriguing promise of enabling ``seeing through walls''. Recent research works have reported such phenomena in different contexts by using several alternative methods, such as propagation through scattering media using speckle correlations\cite{invasive}, scattering-invariant modes \cite{invariantmode}, propagation through lossy materials by complex wave-vector engineering \cite{frezza1,frezza2,frezza3}, propagation through layered Bragg media within forbidden bands using spatial shaping  \cite{photonicrystal}, non-Hermitian exceptional points \cite{exceptpoint}, or exploiting parity-time-symmetry in evanescent waves \cite{paritytimegain}, among others.\\

Metasurfaces are two-dimensional engineered structures \cite{meta1,meta2} whose meta-atoms may control, locally, the amplitude and phase of the transmitted or reflected fields. Linear phase gradient metasurfaces are equivalent to adding a real transverse wave-vector component to a transmitted beam, enabling beam redirection and generalising Snell's Law \cite{capassoscience,flat1,flat2}. We raise the possibility of achieving a complex-valued additive transverse wave-vector to enable greater control. A complex wave-vector corresponds to a linear phase gradient simultaneous with an exponentially decaying amplitude profile. This further generalises Snell's law to complex angles, enabling surprising possibilities.\\

A particularly tough challenge is to achieve transmission of monochromatic light through an opaque medium modelled by a uniform isotropic complex refractive index $n=\sqrt{\varepsilon\mu}=n'+in''$ with a non-vanishing imaginary part. Inside such a medium, the momentum eigenmode solutions to the wave-equation, whose spatial dependence is given by an exponential $e^{i \mathbf{k}\cdot \mathbf{r}}$, exist under the condition that the wave-vector $\mathbf{k}$ must be complex, owing to the Helmholtz equation $\mathbf{k} \cdot \mathbf{k}=k_x^2+k_y^2+k_z^2=(nk_0 )^2$. This implies that $\mathbf{k}=\mathbf{k}'+i\mathbf{k}''$ has simultaneously non-zero real and imaginary vector parts, so the spatial dependence of the wave explicitly becomes $e^{i\mathbf{k}'\cdot\mathbf{r}}\cdot e^{-\mathbf{k}''\cdot \mathbf{r}}$  associated with an unavoidable exponential decay following the direction of the imaginary term $\mathbf{k}''$, corresponding to an inhomogeneous wave. Conventionally, this attenuation profile happens in the direction of penetration inside the material, causing exponential attenuation through the medium. However, this must not always be the case. Maxwell’s equations allow freedom regarding the direction in which waves may be attenuated. Indeed, in \cite{frezza1,frezza2} it was found theoretically and recently confirmed experimentally \cite{frezza3}, that propagation inside an opaque medium exhibits deeply penetrating waves, by tuning the imaginary wave-vector component $\mathbf{k''}$ to be orientated parallel to the incident interface, obtaining only a purely real component of $\mathbf{k}\cdot\hat{\mathbf{u}}$ in the direction $\hat{\mathbf{u}}$ that penetrates the opaque material. This was proposed by illuminating the opaque media from lossless free space with inhomogeneous waves, such as those coming from a leaky waveguide, at carefully engineered angles \cite{frezza1,frezza2}. In this work, we propose achieving the desired wave-vectors using metasurfaces, greatly inspired by its diffraction features, but extending the concept of phase gradients into a phase-and-amplitude profile which effectively translates into complex wave-vector contributions. We also expand the family of solutions that propagate difractionlessly inside an opaque medium, beyond a single inhomogeneous wave, to include the cosine and Bessel beams family.

\begin{figure*}[!htbp]
    \includegraphics[width=1\textwidth]{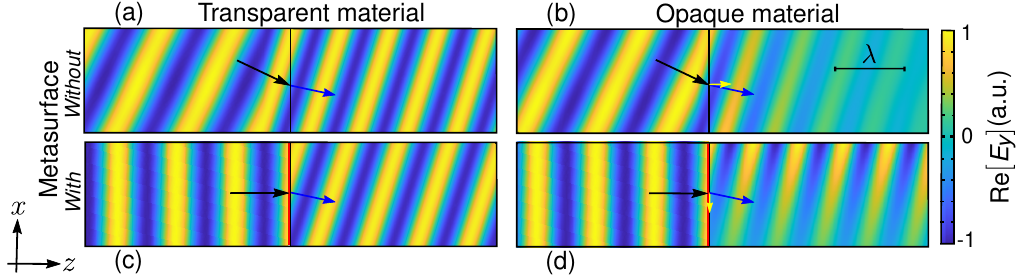}
    \caption{Electric field (real part of TE electric field component) refracted by a smooth interface in transparent (a,b) or opaque (c,d) materials, with (c,d) and without (b,d) a metasurface designed to modify the transmitted wave-vector. The wave incident from free space $n_0=1$ enters a transparent optical material $n_1=1.4$ (a,c) or an opaque optical material (b,d) with effecticve refractive index $n_2=1.4+0.15i$ . In (a), an obliquely incident wave refracts upon entering a transparent material, demonstrating Snell's law and conservation of transverse momentum. Next, in (b) an obliquely incident wave refracts and experiences attenuation upon entering an opaque material. The imaginary component of the wave-vector points along $z$ due to conservation of transverse momentum $k_x \in \mathbb{R}$.Now in (c) a normally incident wave refracts into a transparent material, at the same angle as in (a), thanks to the interaction with a designer phase gradient metasurface with transmission coefficient $t \propto e^{i \Delta k_x x}$ that induces a synthetic transverse wave-vector $\Delta k_x$.  In (d) a normally incident wave interacts with a phase-and-amplitude metasurface with transmission coefficient $t \propto e^{i (\Delta k_x' + i \Delta k_x'') x} = e^{i \Delta k_x' x} e^{-\Delta k_x'' x}$ which synthesises a complex transverse wave-vector $\Delta k_x' + i \Delta k_x''$ (with real and imaginary components, associated to the phase and amplitude of the metasurface, respectively). The metasurface in (d) is designed such that the resulting wave-vector is real along the penetrating $z$ direction, $k_z \in \mathbb{R}$, achieving invariant propagation inside the opaque material. Incident wave-vector (black arrow), real (blue), and imaginary (yellow) components of the transmitted wavenumber are shown as vectors. Colour scale (arbitrary units) is shared for all plots.}
\label{fig:snell}
\end{figure*}

Let’s explore from first principles the limitations and constraints imposed by Maxwell’s equations of electromagnetism inside an opaque material, in a general case. Without loss of generality, we choose our axes such that the axis $z$ is oriented in the direction in which we want to penetrate the material (for instance, normal to the surface of a slab of opaque material that we want to ``see'' through). Can we find waves that deeply propagate into an opaque material, with no exponential attenuation in the desired $z$-direction? The spatial dependence of the wave in that direction is given by $k_z$. We know from the dispersion relation $\mathbf{k}\cdot \mathbf{k}=k_x^2+k_y^2+k_z^2=(nk_0)^2$  that

\begin{equation}\label{eq:kt}
k_z=\left[(nk_0)^2-k_x^2-k_y^2\right]^{1/2}=\left[(nk_0)^2-k_t^2\right]^{1/2},
\end{equation}

\noindent where $k_t$ is often called the transverse wavenumber \cite{pakscience}. The above equation is often expressed in terms of the transmitted angle $\theta$, via $k_t = n k_0 \sin \theta$ and $k_z = n k_0 \cos \theta$. In a lossy or opaque material, $nk_0 = (n' + i n'')k_0$ corresponds to a complex number, which under conventional circumstances suggests that $k_z$ must be complex too, as implied in Eq.~\eqref{eq:kt} and in $k_z = n k_0 \cos \theta$. A complex $k_z$ with non-zero imaginary part gives rise to the attenuation profile characteristic of the Beer-Lambert law in an opaque material. However, this is not necessarily the case. The transverse components $k_x$ and $k_y$ (and hence $k_t$ and $\theta$) may also be complex numbers, corresponding to complex refraction angles. A complex-valued angle $\theta$ can be fine-tuned such that $\cos \theta$ perfectly counterbalances the imaginary part of $nk_0$ to achieve a purely real product $k_z = n k_0 \cos \theta$, and hence a non-attenuating transmission in the $z$-direction. Following Eq. \eqref{eq:kt}, this corresponds to finding a complex $(k_t)^2$ to counterbalance the imaginary part of $(nk_0)^2$. This was precisely the active approach exploited in \cite{frezza1,frezza2}. Here, we propose achieving the required complex $k_t$ by using a passive metasurface.\\


Metasurfaces are an established method to engineer the transverse wave-vector of an impinging wave by means of phase gradients. For this purpose, consider a metasurface on an interface $z=0$; if this metasurface introduces a spatially varying transmission phase gradient $\Phi$ that has a linear dependence in some direction on the $xy$ plane, $\Phi(x,y) =\Delta k_x x+\Delta k_y y$, then the effective transmission coefficient  is $t(x,y)\propto e^{i\mathbf{\Phi}}=e^{i \Delta k_x x+ i \Delta k_y y}$, such that any incident wave $E_{\mathrm{inc}}(z=0) \propto e^{ik_x x+ik_y y}$ will acquire an additional phase upon transmission according to the following expression:

\begin{align} \label{tt}
E_{\mathrm{t}} (z=0) &= t(x,y) E_{\mathrm{inc}} (z=0) \nonumber \\  &\propto t(x,y)e^{ik_x x+ik_y y}\\
&\propto e^{i(k_x+\Delta k_x) x+i(k_y+\Delta k_y) y},\nonumber
\end{align}

\noindent which is mathematically equivalent to a change in the incident transverse wave-vector $(k_x,k_y) \to (k_x+\Delta k_x,k_y+ \Delta k_y)$, i.e. $\bold{k}_t \to \bold{k}_t + \Delta \bold{k}_t$. This method was used as a generalisation of Snell’s law of refraction \cite{capassoscience} and gave rise to many applications of metasurfaces for flat lensing and beam steering \cite{flat1,flat2}.

As is well known, conventional refraction at a smooth boundary between two materials ultimately stems from the conservation of the transverse wave-vector in the interface. By $k_x$ and $k_y$ being conserved between the incident and transmitted waves, we can derive from Eq.~\eqref{eq:kt} that $k_z$ must be genuinely different in each medium, according to their different refractive index $n$, giving rise to the angles of incidence and refraction of the wave, summarised in Snell-Descartes’ law, Fig.~\ref{fig:snell}(a). While this formula is typically used in the context of transparent materials with both real refractive indexes simultaneously, the conservation of the transverse momentum argument always applies. If $n$ is complex in the second medium, $k_x$ and $k_y$ being the transverse components in this interface are still conserved, and hence will be real because the incident wave comes from a transparent medium, and so then $k_z$ will be complex, resulting in an attenuation into the second medium, as shown in Fig.~\ref{fig:snell}(b). The use of a metasurface breaks the requirement of conservation of transverse momentum, by breaking the translational symmetry of the interface. Indeed, a metasurface with a linear phase gradient permits effective additive changes to the transmitted transverse wave-vector as described above, $(k_x,k_y) \to (k_x+\Delta k_x,k_y+ \Delta k_y)$, achieving the corresponding change in angle, generalising Snell’s law \cite{capassoscience}, corresponding to Fig.~\ref{fig:snell}(c).\\

Here, we propose to further generalise the change in transverse wave-vector in the transmitted wave, not limiting ourselves to changing the angle of transmission, but instead including complex values into $\Delta \bold{k}_t$ such that $(k_x,k_y) \to (k_x+\Delta k_x,k_y+ \Delta k_y) = (k_x+\Delta k_x' + i \Delta k_x'',k_y+ \Delta k_y'+i\Delta k_y'')$. In order to achieve this, one only needs the transmission coefficient to be $t\propto e^{i \Delta k_x x + i \Delta k_y y} = e^{i( \Delta k_x' x + \Delta k_y' y)} e^{-(\Delta k_x'' x +\Delta k_y'' y)}$ which represents a simultaneous phase $\mathrm{arg}(t)$ and amplitude $|t|$ control. By doing this, $k_x$ and $k_y$ can be simultaneously engineered by the metasurface at will, to attain any complex value. This is mathematically equivalent to complex-valued angles of refraction, thus representing a complex refraction metasurface. Specifically, the complex $k_x$ and $k_y$ may be engineered in such way to obtain a real-valued $k_z$, even inside a lossy or opaque material, and therefore achieving deep perfect penetration inside the material. Note that the engineering of the transverse wave-vector can be achieved for any angle of incidence, including normal incidence of a plane wave on the interface, as shown in Fig.~\ref{fig:snell}(d).\\

In fact, Fig.~\ref{fig:snell}(d) demonstrates how, by adding an exponential amplitude dependence to the field profile along the $x$ axis, $k_z$ can be made purely real, resulting in transmission with no attenuation along the $z$-axis. Of course, attenuation is still present on this opaque material, but it is happening along the $x$ rather than the $z$-direction into which we want to penetrate. In essence, the use of the ``complex-value'' metasurface allows us to manually select the direction in which the evanescent decay of the amplitude will occur. \\

\section{Analytical derivation}

Emphasis on the complex nature of the wave equation will be the starting point of this derivation; by looking at Eq.~\eqref{eq:kt} we can explicitly find the value that $k_t$ needs to attain in order to achieve a purely real $k_z$ even when $n=n'+in''$ is a complex quantity. For this to happen, two conditions must be simultaneously fulfilled by the argument of the square root in Eq.~\eqref{eq:kt}: firstly, we need that $\mathrm{Im}(k_z^2)=0$, and secondly $\mathrm{Re}(k_z^2 )>0$. These two hyperbolic conditions guarantee a real $k_z$. It is straightforward to show that the first rectangular condition leads to $k_t' k_t''=n' n''$, while the second inequality is equivalent to $k_t'^{2}-k_t''^{2}<k_0^2(n'^{2}-n''^{2})$. Now, the conic solutions for $k_t$ are not unique; there is a full range of values of optimised complex $k_t^{\mathrm{opt}}$ solutions fulfilling these two conditions simultaneously. The family of solutions can be written as an explicit expression with a real parameter $f$ corresponding to a single degree of freedom such that:

\begin{equation}\label{eq:f}
k_t^{\mathrm{opt}}=k_0 \underbrace{\left( n'f+i \frac{n''}{f} \right)}_{n \sin (\theta_{\mathrm{opt}})} \mbox{\phantom{s}  for any  \phantom{s}}  0<\abs{f}<1,
\end{equation}

\noindent where $\theta_{\mathrm{opt}} = \sin^{-1}(k_t^\mathrm{opt}/nk_0)$ represents the complex angle of refraction. Under this finely tuned condition, substituting Eq.~\eqref{eq:f} into Eq.~\eqref{eq:kt}, the value of $k_z$ reduces to a purely real value analytically given by $k_z=\frac{k_0}{\abs{f}}\sqrt{(1-f^2)(n''^{2}+n'^{2}f^2)}$, achieving no exponential decay along $z$, as initially desired. This of course comes at the cost of having a fully complex $k_t$ which requires exponential decay (and increase) of the amplitude in the transverse plane. A metasurface with complex $(\Delta k_x,\Delta k_y)$ can be used to generate the required $k_t$ for any incident wave.\\

\begin{figure*}[!htbp]
    \includegraphics[width=1.00\textwidth]{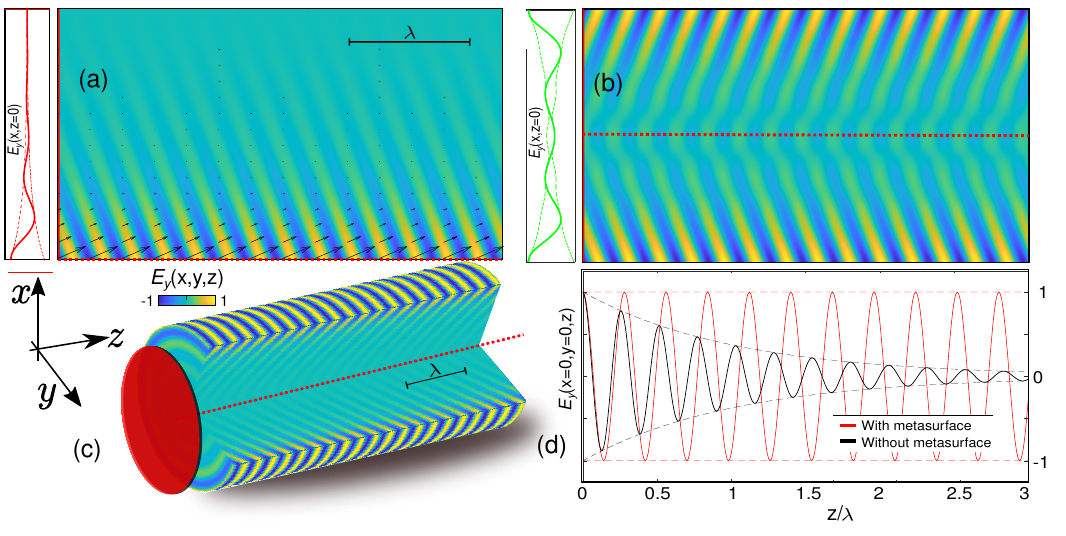}
    \caption{\textbf{Idealised field profiles and perfect penetration}. Electric field evolution inside an opaque material when designed electric field boundary conditions are placed at $z=0$ (i.e. metasurface) for different incident profiles. (a) A complex-$k_x$ plane wave $E_y(x) = e^{i k_t^{\mathrm{opt}} x}$, (b) a complex-argument cosine beam $E_y(x) = \cos(k_t^{\mathrm{opt}} x)$, and (c) a complex-argument Bessel beam $E_y(\rho) = J_0(k_t^{\mathrm{opt}} \rho)$ (see exact derivation in Supplementary). The three are examples of propagation-invariant and difractionless beams through an opaque material with refractive index $n=3.8823+0.01564i$, corresponding to silicon at HeNe laser wavelength $\lambda=633 \mathrm{nm}$ \cite{nsi}. The structured electric field at $z=0$ (which in principle could be synthesized with a metasurface) follows the equations given in the main text with complex transverse wave-vectors, and with fixed parameter $f=0.11$. In boxes, the initial structured profiles $E_{\mathrm{inc}}(z=0)$ real part (continuous line) and amplitude (dashed line) are shown. In (d) the field propagation profile along the $z$ axis is shown, observing a constant amplitude (dashed line) and non-decaying harmonic oscillation (real part, continuous line) in the centre of the transmitted beam, compared to the decaying profile of the case without metasurface (black) when $E_{\mathrm{inc}}(z=0) = 1$ is a normally incident plane wave with no structuring. The real part of the time-averaged Poynting vector is also shown in (a) as black arrows. Colour scale (in a. u.) is shared for all plots.}
\label{fig:infinite}
\end{figure*}

An obvious question that arises when looking at this technique is where does the energy come from, to achieve constant field amplitude along $z$, given that the material may be actively absorbing energy. The answer is to be found by looking at the Poynting vector [black arrows in Fig.~2(a)]. The energy lost in absorption is replaced by the energy coming from the edges where the field profile blows up in amplitude. This is in accordance with Poynting's theorem: any absorption at a point is exactly accounted via a net influx of power into that point. This is analytically proven in the Supplementary Material. At first sight, therefore, it seems that this technique should be impossible to implement in practice, because an exponentially growing diverging amplitude would need to be attained by the metasurface as $x \to \infty$, suggesting that an active metasurface with ever-increasing gain would be required. This problem will later be alleviated using passive metasurfaces in which the required transmission profile is windowed in space, such that the transmission coefficient remains always lower than unity. This windowing process locally and partially preserves the non-decaying effect. Before delving into the windowed case, let’s first try to understand the ideal ``active'' case, in which the metasurface exhibits  an idealized exponentially growing transmission.\\

Let’s begin with a two-dimensional problem, where $k_y=0$, and where a transverse electric polarisation is assumed $\mathbf{E}(\mathbf{r})=E_y (\mathbf{r})  \mathbf{\hat{y}}$. In this case, after interaction with the metasurface, we want to achieve a transverse wavenumber $k_x=k_x^{\mathrm{inc}}+\Delta k_x=k_t^{\mathrm{opt}}$ that matches the optimal condition of Eq.~\eqref{eq:f}, such that $k_z$ is a purely real value inside the opaque material. This can clearly be achieved if a normally incident plane wave in free space ($k_x^{\mathrm{inc}}=0$) of amplitude $E_{\mathrm{inc}}$ interacts with a metasurface whose transmission coefficient equals $t=e^{ik_t^{\mathrm{opt}} x}$ (which will involve both phase and amplitude changes in accordance to the real and imaginary parts of $k_t^{\mathrm{opt}}$). This metasurface will result in a transmitted field profile at $z=0$ equal to $\bold{E}(x,z=0)=\bold{E}_{\mathrm{inc}} t=E_{\mathrm{inc}} e^{ik_t^{\mathrm{opt}}x}\mathbf{\hat{y}}$. This field profile induced at the interface $z=0$ will now propagate inside the opaque material in accordance to Eq.~\eqref{eq:kt}. When propagated along $z$ using the transfer function $e^{ik_z z}$, with now a purely real $k_z$ (as corresponds to $k_t^{\mathrm{opt}}$ by construction), this will result in a transmitted field $\bold{E}(x,z)=E_{\mathrm{inc}} e^{i(k_t^{\mathrm{opt}} x+k_z z)}\mathbf{\hat{y}}$. This is an inhomogeneous plane wave whose amplitude $\abs{\mathbf{E}(\bold{r})}$ does not decay with $z$. This idealised case is illustrated in Fig.~2(a), where for illustration purposes we chose an opaque silicon material at a HeNe laser wavelength. The figure also displays the real Poynting vector as black arrows, showing that the energy indeed comes from the regions of high amplitude at the bottom, in order to sustain the constant amplitude that is achieved along $z$ for every line of constant $x$. \\

Interestingly, a single exponential solution (inhomogeneous plane wave) as shown in Fig.~2(a) is not the only possibility that could exhibit this property. In fact, there are many valid values of $k_t^{\mathrm{opt}}$ that achieve a real value of $k_z$, as described by the free parameter $f$ in Eq.~\eqref{eq:f}, and thanks to linearity of the solutions, any combination of valid exponentials with different $f$ values will still be a penetrating solution with no attenuation along $z$. More simply, however, we can consider by symmetry the combination of two solutions as in Fig.~2(a) but with opposite signs in $k_x=\pm k_t^{\mathrm{opt}}$. Both solutions will have the same real $k_z$ and hence the following field distribution is obtained: $\bold{E}(x,z)=E_{\mathrm{inc}} [e^{i(k_t^{\mathrm{opt}} x+k_z z)}+e^{i(-k_t^{\mathrm{opt}} x+k_z z)}] \bold{\hat{y}} = 2 E_{\mathrm{{inc}}} \cos(k_t^{\mathrm{opt}}x)e^{ik_z z}\bold{\hat{y}}$. This field, invariant in $z$ and thus penetrating, simply requires establishing a cosine-like field profile at $z=0$. In transparent media this is the well-known 2D non-diffracting cosine beam \cite{cosine1, cosine2, cosine3}. In the case of opaque materials, however, we have the intriguing property that $k_t^{\mathrm{opt}}$, and hence the argument of the cosine function, is a complex number. We call this a complex-argument cosine beam. This can be achieved with the metasurface by tuning both the amplitude and phase of the transmission coefficient following $t \propto \cos(k_t^{\mathrm{opt}}x)$. Not surprisingly, like the individual exponential solutions, the cosine profile $\cos(k_t^{\mathrm{opt}} x)$ blows up in amplitude when $x \to \pm\infty$. Considering this idealised case, the cosine beam indeed propagates arbitrarily deep into the opaque material as shown in Fig.~2(b). This time, energy is coming from both top and bottom, to allow for the non-decaying propagation along $z$ in the center. Other relative amplitudes of the two exponentials could be considered, giving rise to sine beams or any other arbitrary function by careful Fourier decomposition.\\

Finally, we consider three-dimensional solutions. We can produce a linear superposition of complex exponentials whose transverse wave-vector $k_x \hat{\bold{x}} + k_y \hat{\bold{y}}$ points in every possible direction on the $xy$-plane, while still maintaining the same transverse wave-number $k_t^\mathrm{opt} = (k_x^2+k_y^2)^{1/2}$ by taking $k_x = k_t^{\mathrm{opt}} \cos(\alpha)$ and $k_y = k_t^{\mathrm{opt}} \sin(\alpha)$, with $\alpha \in [0,2\pi)$. This forms a three-dimensional Bessel beam, but with the peculiarity that $k_t$ is complex, and hence the Bessel functions that describe the field distribution have complex valued arguments in the family $J_m(k_t \rho)$, where $\rho = \sqrt{x^2+y^2}$ is the cylindrical radial coordinate and $J_m$ is the Bessel function of first kind and order $m$. There is freedom on $m$ and on the polarisations of the different plane wave components in 3D, which adds much variety and complexity to the mathematics of the solution, described in detail in the Supplementary Materials; as a simplest example, the electric field corresponding to a $y$-polarised Bessel beam is given by $\bold{E}(\bold{r}) = E_\mathrm{inc} \left[ k_z J_0(k_t^{\mathrm{opt}} \rho)\bold{\hat{y}} - i k_t^{\mathrm{opt}} J_1(k_t^{\mathrm{opt}} \rho) \sin(\phi) \bold{\hat{z}}\right] e^{i k_z z}$, a full solution to Maxwell's equations, and its $y$-component is depicted  in Fig.~2(c). This is indeed the generalisation, to opaque materials, of the well known diffractionless Bessel beams defined in transparent materials. In our case, the argument of the Bessel functions ($k_t \rho$) forming the beam is complex, so the amplitude of the beam grows with $\rho$, but the beam is still invariant with $z$, despite the material being opaque. \\

As we have mentioned before, the obvious caveat of the phenomenon presented here is that an infinite exponential increase in the amplitude of the transmission along the direction parallel to the interface is needed, thus requiring an infinite energy to maintain the propagated beam in a physical realization. To avoid this unphysical scenario, we next study the effect of spatially windowing the initial excitation, with a hope of still achieving non-decaying transmission far from the edges, thanks to the locality of Maxwell equations. This is discussed in the following section.

\section{Finite case: Bullseye metasurface}

Let us avoid the infinitely growing tails of the complex-argument cosine and Bessel beams by windowing the initial field profile at $z=0$ using the following expression:

\begin{equation}\label{eq:windowed}
\bold{E}_{\mathrm{windowed}}(x,z=0)=E_{\mathrm{inc}} \frac{w(x)\bold{E}_{\mathrm{ideal}}(x,z=0)}{\mathrm{max}(\abs{w(x) \bold{E}_{\mathrm{ideal}})})},
\end{equation}

\noindent where $\bold{E}_{\mathrm{ideal}}(x,z=0)$ corresponds to any of the exponentially growing profiles shown in the previous section (Fig.~2), $w(x)$ is a mathematical window function which is vanishing for any $\abs{x}>L/2$, such as a rectangular step window, and we normalize the entire field profile by $\mathrm{max}(\abs{w(x)\bold{E}_{\mathrm{ideal}}})$ such that at no point is the amplitude of $\bold{E}_{\mathrm{windowed}}(x,z=0)$ greater than $E_{\mathrm{inc}}$. This will correspond to a passive metasurface transmission coefficient $\abs{t}<1$, avoiding the need for active amplification. It is reasonable to expect that the results of the windowed case should asymptotically approach the idealised perfect penetration as the window grows $L\to \infty$ and therefore it is interesting to consider how well will the penetration rate work for realistic window sizes. Of course, this windowed profile is no longer a perfect inhomogeneous wave, so we can no longer use simple analytical solutions, and we cannot expect the field inside the opaque material to be translationally invariant along the $z$-direction. However, because Maxwell's equations are local, one can expect that, locally, the windowed version will be similar to the ideal version $\bold{E}_{\mathrm{windowed}} \simeq (\mathrm{max}(\abs{w(x)  \bold{E}_{\mathrm{ideal}}}))^{-1} \bold{E}_{\mathrm{ideal}}$ as $w(x)\sim 1$ at least in a neighbourhood around $(x,z)=0$, i.e. near the central axis of the beam, far from the window edges, corresponding to the bullseye proposal. Like a bull's-eye window in the hull of a ship, this metasurface opens up a small region of transmission on an opaque slab.\\

\begin{figure}[!htbp]
 \centering
    \includegraphics[width=8.6cm]{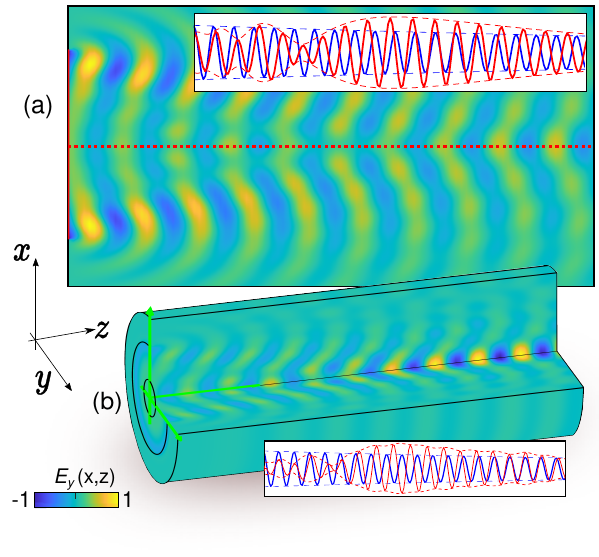}
    \caption{\textbf{Bullseye metasurface: windowed field profiles}. Electric field propagation inside an opaque material when designed boundary conditions (metasurface) are optimised and specified at $z=0$ corresponding to the windowed versions (window length $L=0.654\lambda$) of (a) a complex-argument cosine beam, and (b) a complex-argument Bessel beam. The lossy material corresponds again to silicon at HeNe wavelengths. The internal degree of freedom, set to $f=0.117$, is optimised to obtain the maximum intensity enhancement inside the opaque material compared to an exponential attenuation (no metasurface). The insets shows the field profile for the central position along the $z$ axis as the field propagates inside the opaque material, for the structured windowed cosine beam, i.e. with metasurface (red), and for a non-structured normally incident plane wave, i.e. without metasurface (blue). We find a non-trivial increase in the amplitude of electric field for a wide range of penetration distances that can be effectively tuned for further applications.
    }
\label{fig:finite}
\end{figure}

To study this, we numerically propagate the windowed profile through the opaque medium using a Fourier propagator approach by performing the following steps: we numerically compute the Fourier transform of the field at the plane $z=0$, followed by an $e^{i k_z z}$ transfer propagator function across the $z$ planes, and finally an inverse Fourier transform on each $z$ plane. The resulting fields are shown in Fig.~\ref{fig:finite}, where the intensity profile imposed at $z=0$ is obtained by windowing the ideal cases of cosine and Bessel beam that were previously shown in Figs.~2(b-c). The window size $L$ was optimised in each case for higher amplitudes along $x=0$, looking for a trade-off between $L$ being too-small (and thus no opportunity of resembling the original ideal case) or being too-wide (and thus having to normalise the entire profile by a number that is too big, to keep the exponentially growing tails under control). Remarkably, and with no intentionality on our part, the resulting profile shows a focusing behaviour, resulting in enhanced field amplitude penetrating in the $z$-direction along the line $x=0$, but with low penetration depth elsewhere. This focusing behavior matches well with the fact that the Poynting vector is flowing from the high amplitude edges of the field profile toward the lower amplitude center, at a certain angle, as seen in the ideal cases earlier. Interestingly, it is  possible to directionally control this focusing-like behaviour of the beam inside the material thanks to the internal degree of freedom $f$, inherited from Eq.~\eqref{eq:f}. Note also that changing the size of the window can also lead to a fine tuning of the profile shape inside the material. If this windowed profile can be set up by a metasurface fabricated at the interface of the opaque material, we may achieve, locally, an enhanced transmission through the material (in our examples from Fig.~\ref{fig:finite} we show up to $76\%$ improved intensity of the field at some points inside the material, with optimization of the internal degree of freedom $f$ and the window size $L$) for feasible technological applications.\\

\section{Discussion and future outlook}
A transmitted wave propagating through an opaque material was analytically proven to result from a complex refraction metasurface. This corresponds to a metasurface that generalises Snell's law to complex angles, by using simultaneous phase and amplitude control. The main physical explanation  is that one can swap the direction of decay from a longitudinal to a transverse direction. This suggests novel methods for transmission through opaque media, moving away from active sources and continuous engineering beam shaping, to a more sustainable approach based on a passive structure metasurface. To solve the nonphysical requirement of infinite energy tails, a solution is presented by windowing the ideal case, corresponding to our proposal of a bullseye metasurface. Having proposed the concept of complex refraction bullseye metasurfaces, further work could look at interesting applications like an in-situ wall-mounted metasurface for increased wireless signal transmission through domestic walls or, thanks to the electromagnetic-acoustic formalism analogy, increased sound transmission through ``acoustically opaque'' (dampening) materials. Note that the phenomenon works for any complex refractive index, whether it is associated to losses, or to lossless but still opaque materials, such as a plasma with a negative but real electric permittivity resulting in purely imaginary refractive index. Speculative proposals can therefore be pictured in the context of improved transmission across metals. Medical applications could include the analysis and treatment of malignant cells subcutaneously without damaging the external layer of skin in the human body, without (over)heating of the interface boundaries, with a proper combination of bullseye metasurface profiles. Further work for complex refraction metasurfaces is not limited to achieving a zero imaginary part in $k_z$, but even a negative imaginary part corresponding to exponential increase inside the material, at the cost of even stronger transverse amplitude variation.


\section*{Acknowledgements}
S. P.-P. thanks V. Arribas-López for the rendering of the 3D graphic implementation, A. Díaz-Nadales for the helpful suggestion about the Feynman trick and A. Ortega-Arroyo for the fruitful discussion about the potential applications in Health Sciences. F. J. R.-F. thanks N. Engheta for useful discussions. This work was supported by European Research Council Starting Grant ERC-2016-STG-714151-PSINFONI and EPSRC-2020-20004058-11982 (UK). The authors declare no competing financial interest. 

\bibliography{main}
\appendix

\section{Poynting's theorem in the idealised solution}
At first glance, having a propagation-invariant amplitude inside a lossy material seems impossible from elemental energy conservation arguments. However, the tilt angle present in the Poynting vector solves this puzzle. A constant flow of energy from the high intensity edges compensates for the absorption losses due to absorption. This is always true thanks to the Poynting theorem for monochromatic fields $A(\bold{r}) = -\nabla \cdot \bold{S}$, which states that any energy lost in absorption in a given volume must be accompanied by a net inward flow of power through its boundary. 

Let's check this analytically for the idealised ``infinite metasurface'' solutions shown in Fig.~\ref{fig:infinite}. Consider the electric and magnetic field solutions given in phasor notation as $\bold{E} = \bold{E}_0 e^{i \bold{k}\cdot\bold{r}}$ and $\bold{H} = \bold{H}_0 e^{i \bold{k}\cdot\bold{r}}$, with $\bold{H}_0 = (1/\omega \mu_0 \mu_r)(\bold{k}\times\bold{E}_0)$ and $\bold{k}\cdot\bold{k} = (nk_0)^2$ from Maxwell's equations. First, we calculate the absorption $A(\bold{r})$ on the material. Second, we can calculate the divergence of the Poynting vector at each point. As expected from Poynting's theorem, both are equal.
Inside a homogeneous, isotropic, non-magnetic, linear material, the polarization density is given by $\bold{P}=\varepsilon_0(\varepsilon_r-1)\bold{E} = \varepsilon_0(\varepsilon_r'+i \varepsilon_r''-1) \bold{E}$. This  is responsible for a bound current $\bold{J}=-i\omega \bold{P}$. The absorption at each point in the dielectric can be calculated then as:
\begin{align}
A&=\frac{1}{2}\Re[\bold{J}\cdot \bold{E^*}]\nonumber\\
&=\frac{1}{2}\Im \left[ \omega \varepsilon_0 (\varepsilon_r-1) \| \bold{E} \|^2 \right] \nonumber\\
&=\frac{\varepsilon_0 \varepsilon_r'' \omega\|\bold{E}_0\|^2}{2}e^{-2\bold{k}''\cdot\bold{r}},\nonumber
\end{align}
\noindent where we assumed that $\omega$ is purely real. On the other hand, the time averaged Poynting vector can be obtained as 

\begin{align} 
\bold{S}&=\frac{1}{2}\Re\left[\bold{E}\cross \bold{H}^*\right]\nonumber\\
&=\frac{1}{2}\Re\left[\frac{\bold{E}_0\cross (\bold{k} \cross \bold{E}_0)^*}{\omega\mu_0\mu_r^*}\right]e^{-2\bold{k}''\cdot\bold{r}}\nonumber\\
&=\frac{1}{2}\Re\left[\frac{ \|\bold{E}_0\|^2 \bold{k}^* - \bold{k}^* \cross (\bold{E}_0^*\cross \bold{E}_0)}{\omega\mu_0\mu_r^*}\right]e^{-2\bold{k}''\cdot\bold{r}},\nonumber
\end{align}

\noindent where in the last step we have used both Jacobi identity and Lagrange formula for triple cross products, as well as Gauss' law $\div \bold{E}_0 = i \bold{k}\cdot\bold{E}_0 = 0$ to remove one of the terms. Hence, the divergence of the Poynting vector is:

\begin{align}
\div \bold{S}&=\Re\left[-\bold{k}''\cdot\frac{ \|\bold{E}_0\|^2 \bold{k}^* - \bold{k}^* \cross (\bold{E}_0^*\cross \bold{E}_0)}{\omega\mu_0\mu_r^*}\right]e^{-2\bold{k}''\cdot\bold{r}},\nonumber 
\end{align}

\noindent which for a non-magnetic medium ($\mu_r=\mu_r^*=1$) turns into:

\begin{align}
\label{S3}
\div \bold{S} &= -\frac{1}{\omega \mu_0}\left( \bold{k}' \cdot \bold{k}'' \right) \|\bold{E}_0\|^2 e^{-2\bold{k}''\cdot\bold{r}},\nonumber
\end{align}

\noindent where we applied $\Re[\bold{k}'' \cdot \left[ \bold{k}^* \times \left( \bold{E}_0^*\cross \bold{E}_0 \right) \right]] = (\bold{k}'' \times \bold{k}')\cdot \Re[\bold{E}_0^*\cross \bold{E}_0] = 0$. We also know that $\bold{k}'\cdot \bold{k}'' = \frac{1}{2} \Im[\bold{k}\cdot{\bold{k}}] = \frac{1}{2} \Im[(nk_0)^2] = \frac{1}{2} \omega^2 \mu_0 \varepsilon_0 \varepsilon_r''$. Substituting this into the divergence of the Poynting vector we immediately recover the same expression as the absorption above,  obtaining Poynting's Theorem, as expected \cite{jackson}. This proof holds for any complex $\bold{k}$, as used in the paper.
\\


\section{Vectorial Bessel Beams\\ inside opaque materials}

In this section we will show the analytical form of full vectorial Bessel beams in opaque media which, with suitably selected transverse wave-vectors $k_{t_0}$ being a complex number, form non-diffracting and non-decaying wave solutions in three dimensions.

Consider the general solution to Maxwell's equations in an arbitrary linear, homogeneous and isotropic medium, but potentially opaque, whose electric field can be seen as a linear combination of plane-wave components \cite{wolf,novotny}:

\begin{align}
\bold{E}(\bold{r})&=\int \bold{E}(\bold{k})e^{i \bold{k} \cdot \bold{r}}d\bold{k}\nonumber \\
&=\iint {\bold{E}(k_x,k_y)e^{ik_z z}}e^{i[k_x x + k_y y]}dk_xdk_y,
\end{align}

\noindent where $\bold{E}(k_x,k_y)$ corresponds to the Fourier transform of the electric field $\bold{E}(x,y)$ at the plane $z=0$. The integral is two dimensional because the wave-vector $\bold{k}$ components are constrained by the Helmholtz condition [$\bold{k}\cdot\bold{k} = (nk_0)^2$] therefore we can take $k_z$ as a dependent variable given by $k_z=\sqrt{(nk_0)^2-k_t^2}$ with $k_t = \sqrt{k_x^2 + k_y^2}$. This means that knowing $\bold{E}(k_x,k_y)$ at $z=0$ we can ``propagate" the fields to every $z$ plane simply by using the propagator $e^{i k_z z}$ function.

A Bessel beam is a wave solution whose plane wave constituents all have a constant value of transverse wave-vector $k_t = k_{t_0}$, and hence a single value of $k_z$ given by $k_{z_0}=\sqrt{(nk_0)^2-k_{t_0}^2}$. The angular spectrum of these beams forms a ring in the $(k_x,k_y)$ spectral domain, with fixed radius $k_{t0}$. For a Bessel beam, we can therefore rewrite the angular spectrum as $\bold{E}(k_x,k_y) dk_x dk_y = \bold{E}(k_t,\alpha) k_t dk_t d\alpha = \bold{E}(\alpha) \delta(k_t - k_{t_0}) k_t dk_t d\alpha$, where $k_t$ and $\alpha$ correspond to spectral polar coordinates with $k_x = k_t \cos{\alpha}$ and $k_y = k_t \sin{\alpha}$. The Dirac delta sieving property means we can evaluate the integral along $k_t$ by simply setting $k_t=k_{t_0}$ and we are only left with  an integral along $\alpha$, i.e., along the ring in $(k_x,k_y)$ space:

\begin{align}
\label{eq:genBessel} 
\bold{E}(\bold{r})&= k_{t_0} e^{iz k_{z_0} } \int_0^{2\pi} \bold{E}(\alpha) e^{ik_{t_0}[\cos(\alpha) x +\sin(\alpha) y]} d\alpha.
\end{align}

Notice that the $e^{i k_{z_0} z}$ propagator could be taken outside the integral because $k_{z_0}$ is constant on a Bessel beam, showing that, as long as $k_{z0}$ is a real number, the beam is invariant to propagation in the z-direction apart from acquiring a net phase $k_{z_0} z$. The above calculation is a well known feature of conventional Bessel beams, but we stress here that the derivation at no point assumed that any of the quantities $n$ or $k_{t}$ be real. All conclusions remain true for Bessel beams in opaque materials. For simplicity we will use $k_t$ and $k_z$ in place of $k_{t_0}$ and $k_{z_0}$, with the understanding that in a Bessel beam both are constant.

Next, we would like to evaluate the angular integral in order to write the analytical form of the spatial distribution $\bold{E}(\bold{r})$ in terms of Bessel functions. There are many possible solutions, as we have freedom of choice of amplitude and polarisation distribution $\bold{E}(\alpha)$ around the ring in k-space. The spectrum $\bold{E}(\alpha)$ is a vector quantity, so, in general, we may expand it on a spherical basis of unit vectors in k-space $\{\mathbf{\hat{e}}_k,\mathbf{\hat{e}}_p,\mathbf{\hat{e}}_s \}$ with $\mathbf{\hat{e}}_k$ being the radial unit vector, $\mathbf{\hat{e}}_s$ the azimuthal angle unit vector, and $\mathbf{\hat{e}}_p$ the polar angle unit vector. This basis is orthonormal in the sense $\bold{\hat{e}}_i \cdot \bold{\hat{e}}_j = \delta_{ij} $. The unit vectors depend on $(k_x,k_y)$ and hence on $\alpha$ and $k_t$:

\begin{align}
\mathbf{\hat{e}}_k &= \bold{k}/k \nonumber \\
\mathbf{\hat{e}}_s &= \frac{\bold{\hat{z}}\times\bold{k}}{k_t} = \frac{1}{k_t}\begin{pmatrix}-k_y\\k_x\\0\\\end{pmatrix} = \begin{pmatrix}-\sin(\alpha)\\ \phantom{-}\cos(\alpha) \\0\\\end{pmatrix} \nonumber \\
\mathbf{\hat{e}}_p &= \mathbf{\hat{e}}_s \times \mathbf{\hat{e}}_k = \frac{1}{k k_t} \begin{pmatrix}k_x k_z \\ k_y k_z \\ -k_t^2 \end{pmatrix} = \frac{1}{k}\begin{pmatrix} k_z \cos(\alpha) \\ k_z \sin(\alpha) \\ -k_t \end{pmatrix}. \nonumber
\end{align}
We are free to choose any $\bold{E}(\alpha)$ field as long as it complies with the transversality condition from Gauss' law $\nabla \cdot \bold{E} = 0$ (which becomes $\bold{k}\cdot\bold{E}(\alpha)=0$). This means that the $\bold{\hat{e}}_k$ component of $\bold{E}(\alpha)$ must be zero. Therefore, we are left, in the most general case, with

\begin{align}
\bold{E}(\alpha) = E_p(\alpha) \bold{\hat{e}}_p(\alpha) + E_s(\alpha) \bold{\hat{e}}_s(\alpha) \nonumber
\end{align}

\noindent and we can choose any $2\pi$-periodic scalar functions for $E_s(\alpha)$ and $E_p(\alpha)$. Substituting that into Eq.\ (\ref{eq:genBessel}) yields the most general formulation for a Bessel beam.

Let's now specify some simple particular case. For example, if we wish to have a linearly polarised Bessel beam (with no $x$-component) we can choose the following spectra $[E_p,E_s] \propto E_0[\sin(\alpha), \frac{k_z}{k} \cos(\alpha)]$. Under this condition, the vector spectrum becomes:

\begin{align}
\bold{E}_1(\alpha) &= E_0 \left[ \sin(\alpha) \bold{\hat{e}}_p(\alpha) + \frac{k_z}{k} \cos(\alpha) \bold{\hat{e}}_s(\alpha) \right] \nonumber \\ 
&= \frac{E_0}{k} \begin{pmatrix}0 \\k_z \\ -k_t \sin(\alpha)\end{pmatrix}
\end{align}

\noindent where we substituted the unit vectors defined above. This spectrum can now be substituted into Eq.\ (\ref{eq:genBessel}):

\begin{align}
\label{eq:BesselIntegral1}
\bold{E}_1(\bold{r}) &= E_0\frac{k_t}{k}e^{ik_z z}   \nonumber \\
&\ \ \ \ \ \int_0^{2 \pi}
    \begin{pmatrix}
    0\\
    \phantom{-}k_z\\
    -k_t \sin(\alpha)\\
    \end{pmatrix}
     e^{ik_t[\cos(\alpha) x +\sin(\alpha) y]}d\alpha,
\end{align}

\noindent and with the substitution $\sin(\alpha) = (e^{i\alpha} - e^{-i\alpha})/2i$, this integral can be evaluated analytically using the Hansel-Bessel formula, which in general takes the form:

\begin{align}
\int_0^{2\pi}&e^{i m\alpha}e^{i k_t [\cos(\alpha)x+\sin(\alpha)y]}d\alpha \nonumber \\
&= \int_0^{2\pi}e^{i m\alpha}e^{i k_t \rho \cos(\phi-\alpha)}d(\phi-\alpha) \nonumber \\
&= 2\pi e^{i m(\phi + \pi/2)}J_m(k_t \rho),
\end{align}

\noindent with $m$ an integer constant, $J_m(k_t\rho)$ a Bessel function of the first kind of order $m$, and $\{\rho,\phi,z\}$ corresponding to canonical cylindrical coordinates in real space such that $x = \rho \cos(\phi)$, $y = \rho \sin(\phi)$. Applying this formula, Eq.\ (\ref{eq:BesselIntegral1}) can be analytically integrated to arrive at a closed-form solution:

\begin{align}
\bold{E}_1(\bold{r}) &=  E_0 \frac{2\pi}{k}  k_t
    \begin{pmatrix}
    0\\
    \phantom{i}k_z\cdot J_0(k_t\rho) \phantom{\sin\phi}\\
    - i k_t\cdot J_1(k_t\rho) \sin(\phi)
    \end{pmatrix}
e^{i k_z z}.
\end{align}
This is a full exact solution of Maxwell's equations, solving both the wave equation and Gauss' law, representing a linearly polarised Bessel beam: the analytical expression matches exactly that of a well-known Bessel beam on a transparent material, but here we are proposing that $k_t$ may be a complex number, without affecting the derivation, and judiciously designed in accordance to Eq.~\ref{eq:f} such that $k_z$ is real even inside an opaque medium with complex refractive index $n$, therefore achieving a non-diffracting and non-decaying Bessel beam inside it. The $y$-component of this field is shown on Fig.~\ref{fig:infinite}(c). A numerical implementation with finite window corresponds to Fig.~\ref{fig:finite}(b) in the main text.
\end{document}